\documentclass{article}
\usepackage{graphicx}
\usepackage{color}
\usepackage{authblk}
\usepackage [english]{babel}
\usepackage [autostyle, english = american]{csquotes}
\usepackage{hyperref}
\MakeOuterQuote{"}

\begin{document}
\title{{\bf Ensemble Fluid Simulations on Quantum Computers}}

\author[1]{Sauro Succi\thanks{\href{mailto:sauro.succi@gmail.com}{sauro.succi@gmail.com}}}
\author[2]{Wael Itani\thanks{\href{mailto:itani@nyu.edu}itani@nyu.edu}}
\author[2,3,4]{Katepalli R. Sreenivasan\thanks{\href{mailto:katepalli.sreenivasan@nyu.edu}{katepalli.sreenivasan@nyu.edu}}}
\author[5]{Ren\'e Steijl\thanks{\href{mailto:rene.steijl@glasgow.ac.uk}{rene.steijl@glasgow.ac.uk}}}
\affil[1]{Fondazione Istituto Italiano di Tecnologia
\authorcr
Center for Life Nano-Neuroscience at la Sapienza
\authorcr
Viale Regina Elena 291, 00161 Roma, Italy}
\affil[2]{Tandon School of Engineering, New York University
\authorcr Brooklyn, New York, NY 11201 , United States of America}
\affil[3]{Courant Institute of Mathematical Sciences, New York University
\authorcr New York, NY 10012, United States of America}
\affil[4]{Department of Physics, New York University
\authorcr
New York, NY 10003, United States of America}
\affil[5]{James Watt School of Engineering,
University of Glasgow
\authorcr
G12 8QQ Glasgow,United Kingdom}
\maketitle
\begin{abstract}
We discuss the viability of ensemble simulations of fluid flows 
on quantum computers.  The basic idea is to formulate a functional  
Liouville equation for the probability distribution 
of the flow field configuration and recognize that, due to its linearity,  
such an equation is in principle more amenable to quantum 
computing than the equations of fluid motion.
After suitable marginalization and associated closure, the
Liouville approach is shown to require several hundreds
of logical qubits, hence calling for a major thrust in current
noise correction and mitigation techniques.

\end{abstract}


\section{Introduction}

The extreme complexity of most problems in modern science and 
society poses a very steep challenge to our best theoretical and
computational methods. As an example, even the most powerful supercomputers,
reaching up to exascale operations (one billion billions floating point operations
per second) pale in front of the task of predicting the
weather on the planetary scale based on the direct simulation of the
equations of fluid motion \cite{PALMER}.
Besides, this and similar problems are typically subject to various sources
of uncertainty arising from the initial data and other parameters affecting the solution. 
As a result, each single case-study requires several realizations in order to accumulate
sufficient statistical information (Ensemble Simulations), further reinforcing
the quest of computational power.  

Given that electronic computers are facing very stringent energy 
constraints, alternative simulation strategies are constantly sought.
Among these, enormous efforts have been devoted in the last decade towards
the development of quantum computers, using hardware devices capable of
exploiting the ability of quantum systems to occupy a multitude of 
states at the same time (quantum entanglement).
The immediate advantage is that a quantum system can {\it in principle} 
perform a multitude of parallel quantum computations, as opposed to classical
computers which can only operate on binary states (bits).
Lately, not a day goes by without hearing the last quantum computing 
breakthrough. However, leaving aside the hype \cite{HYPE}, the fact remains
that turning the immense potential of quantum computing into a concrete
tool for scientific purposes remains very challenging. 
The reasons are many, but, in a nutshell, entanglement
is very fragile and tends to crumble pretty quickly under the effects 
of environmental noise, which is extremely hard to avoid at any reasonable
temperature---a problem know as ``decoherence''.
Notwithstanding these major barriers, it is worth exploring the contributions that quantum computers
can possibly make to the prospect of ensemble simulations of fluid flows.

\section{Ensemble Simulations}

Ensemble simulations have gained popularity in the recent years, thanks 
to the availability of large supercomputers. 
The main idea is accumulate statistics over the many sources 
of uncertainties that are associated, for instance 
with weather forecasting, by running series of simulations with different initial
conditions and/or parametric realizations \cite{NAVAR,PALMER}. 

To illustrate the idea we consider a set of nonlinear partial differential equations
and discretize them on a grid with, say, $G$ grid points.
Let $\vec{u}(t)$ be the set of unknowns after discretization, for instance the
three-dimensional velocity field of a fluid flow; they obey
a set of $O(G)$ first-order ODE's in (generalized) Langevin form, given by
\begin{equation}
\label{LANGE}
\frac{d\vec{u}}{dt} = f(\vec{u};\vec{\lambda}),
\end{equation}
with initial conditions $\vec{u}(0)=\vec{u}_0$.
In the above, $\vec{\lambda}$ stands for a set of parameters subject 
to various sources of uncertainty, thus acting like ``noise'' on the system.
Ensemble simulations correspond to the generation of statistics
of solutions upon changing initial conditions and/or perturbing the
system parameters.
Formally,  this amounts to generating a probability 
distribution function (PDF) for the solutions $\vec{u}(x,t)$,  defined by:
\begin{equation}
\label{PDF}
p(\vec{u},t) |\delta \vec{u}| = \frac{\delta t}{T},
\end{equation}
where $\delta t$ is the time spent by the set of trajectories, spanning the
time interval $[0 \le t \le T]$, in a volume of phase-space $|\delta \vec{u}|$.
Generating the trajectories is extremely demanding, since by construction
each single simulation is set to stress the most advanced 
computer resources to their limit \cite{NAVAR,ENSIM}.

Quantum computing could help realizing an exponential speedup
on each of these simulations. However, besides all standard concerns affecting
quantum computing, two additional issues stand on the way of this program: 
{\it quantum mechanics is linear and unitary, while the physics 
of fluids is neither} Even when energy is explicitly conserved, the fluid model would not be norm-preserving, thus non-unitary.
\footnote{We make abstraction of inviscid fluids, which are nonetheless a (useful) idealization }.

Several ways around these problems 
are currently under exploration, based on various 
strategies, some of which resort to Carleman linearization 
of the fluid equations \cite{CHILDS,MEZZA,WAEL}, while
some others leverage nonlinear quantum ODE solvers \cite{GAITAN}.  In computational fluid dynamics applications of quantum computers, the use of a hybrid quantum/classical formulation is the most widely used approach to deal with the non-linearity of the governing equations of fluid dynamics \cite{GAITAN, RENE,Sachin}, effectively by accounting for this in the classical part of the algorithm. All of these, however, focus on the solution of the dynamic
equations of motion, with no focus on ensemble simulations.

In this brief note, we sketch a potential strategy which 
offers two major assets at the outset. 
First, it captures by construction all the statistical information that is sought
on the system dynamics (statistical dynamics).
Second, it does not resort to any linearization of the dynamic 
equations, but starts directly from an inherently linear 
representation of the corresponding probability distribution function (PDF). 

The passage from Newtonian dynamics to statistical dynamics is a
standard topic in statistical physics, where it is known as
Liouville formulation of classical N-body mechanics. This formalism is elegant and conducive to very valuable approximations,
mostly at the level of one-body effective kinetic equations, the most outstanding
examples of which are the Boltzmann and Fokker-Planck equations.

Unfortunately, at least on classical computers, working with Louiville equations is completely unviable
since the N-body distribution function lives
in a $O(N)$-dimensional space, with $N$ of the order of
the number of grid points of the dynamic simulation,
hence easily in the order of billions or more for current 
supercomputer simulations. 

This looks like a ``medicine-is-worse-than-the-disease" scenario, and it
is therefore of interest to explore what quantum computing
could possibly contribute to easing up the difficulty.

\section{Functional Liouville Equation}

By virtue of the Liouville theorem (Fig 1), the N-point
PDF associated with the (nonlinear) Langevin equations obeys a 
linear Liouville Fokker-Planck kinetic equation (LFPE) of the form
 \begin{equation}
\label{LFPE}
\partial_t p_N + \sum_{i=1}^N \partial_{u_i} [f(u)p_N-D \partial_{u_i} p_N] = 0,
\end{equation}
where $D$ is the diffusion coefficient associated with the noise
in the (linear) Langevin equation (\ref{LANGE}).
\footnote{
Caveat: Since diffusion results from linearising $f(u;\lambda)$ around a 
reference value $\lambda_0$, the corresponding diffusion coefficient 
is generally dependent on the actual flow field, that is, $D=D(u)$.
}

Since the Liouville equation is linear by construction, it can operate
under the same framework as quantum mechanics and in particular, it can benefit
of quantum linear-algebra solvers \cite{QLAS}.

\begin{figure}
\centering
\includegraphics[scale=0.3]{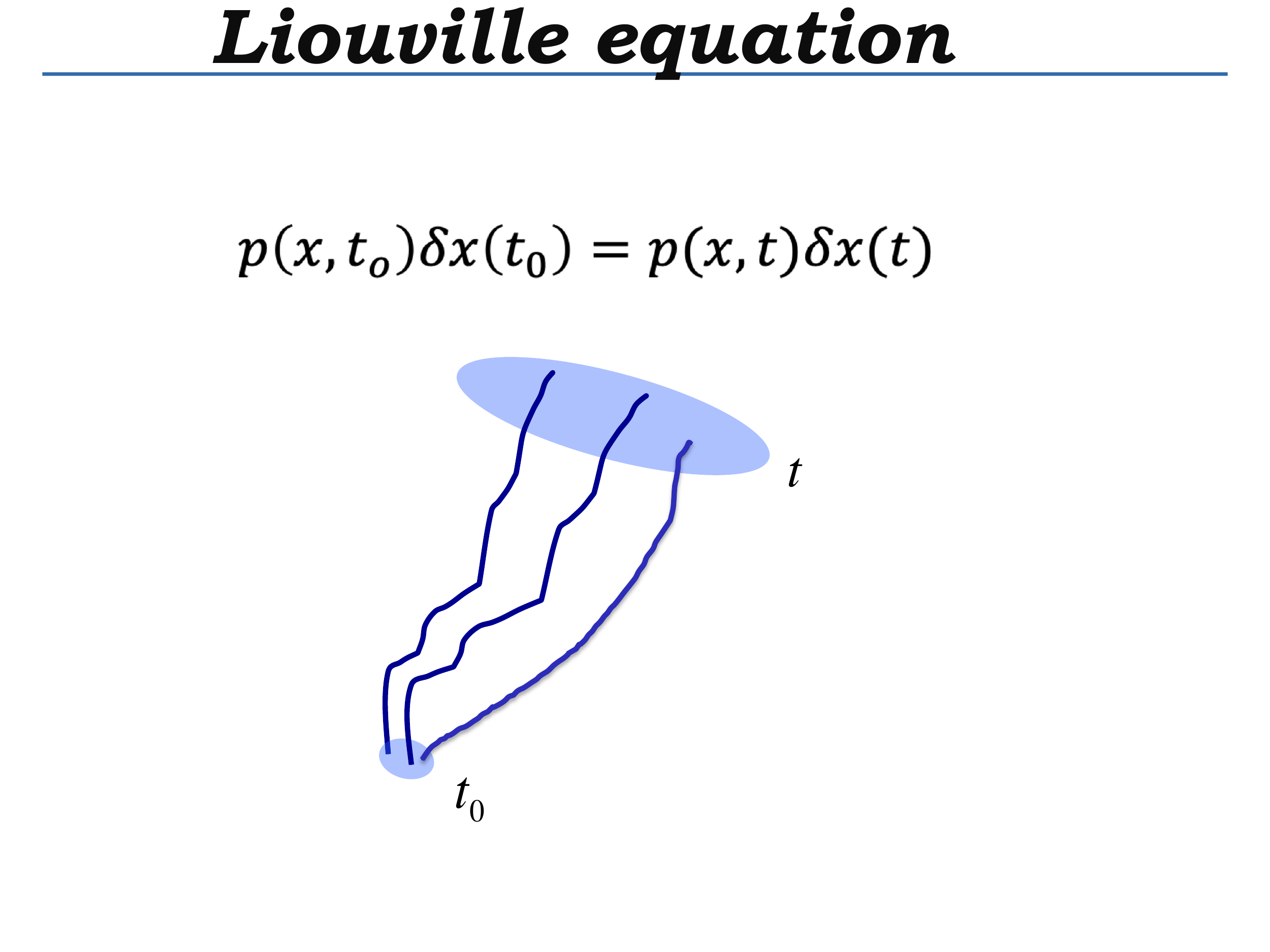}
\caption{Geometrical interpretation of the Liouville equation.
The cloud of points representing various realisations of the system at $t=t_0$ evolves each along its own trajectory dictated by the dynamic equation $\dot{\vec{u}} = f(\vec{u})$, with initial conditions $\vec{u}(t_0)=\vec{u}_0$. As time unfolds, the cloud changes its shape but not its volume (if $\nabla \cdot \vec{u} \ne 0$) and consequently the  probability distribution  $p(\vec{u},t)$ is invariant along the trajectory $\frac{dp}{dt}=0$, leading to the Liouville equation.}
\end{figure}

\subsection{Taming the Dimensional Curse}

The Liouville equation is very elegant but operationally unfeasible, since it
lives in a ultra-dimensional space with as many dimensions as the number
of grid sites where the direct simulations are performed; as already stated, this number is easily in excess
of many billions for present-day supercomputers.  
This is the so called dimensional curse, affecting many problems in
modern science and engineering.

The main merit of the Liouville equation, though, is that it opens up lower-dimensional
approximations which can often capture the essence of the physical
problem at hand. The technical procedure is called
{\it marginalization} and consists of deriving equations for 
lower-order marginals of the original N-body PDF.
Formally, this is obtained by projecting out the unwanted/unnecessary variables
by integrating them out, as follows:
\begin{equation}
P_M(u_1 \dots u_M) = \int P_N(u_1 \dots u_N) du_{M+1} \dots du_N,
\end{equation}
where $1 \le M \leq N$ defines the order of the marginal distribution.
By applying the above definition to the N-body
Liouville equation (\ref{LFPE}), one readily obtains
\begin{equation}
\partial_t p_M + \sum_{i=1}^M \partial_{u_i} 
[F(u)p_M-D \partial_{u_i} p_M] = 0,
\end{equation}
where 
\begin{equation}
F(u) \equiv F(u_1 \dots u_M) = 
\int f(u_1 \dots u_N) p_N (u_1 \dots u_N) du_{M+1} \dots du_{N}
\end{equation}
is the effective $M$-body force.
Here, we have assumed no-flux boundary conditions.
From the above relation, it is immediately clear that
the explicit expression of $F(u)$ generally depends on
the unknown N-body PDF, signaling a much expected closure
problem. This is, of course, a key issue for the success of the whole program,
but in the following we shall proceed by assuming that a plausible
closure can be found. In light of the major advances in statistical mechanics
and the theory of coarse-graining this is, after all, a plausible assumption (though its precise realisation has often defined major efforts). 

Next, let us consider a generic observable $A(u_1 \dots u_N)$, whose 
average value is given by
\begin{equation}
\langle A\rangle (t) = \frac{1}{Z(t)} \int_{-\infty}^{+\infty} p(u,t) A(u) du,  
\end{equation}
where
$Z(t)=\int_{-\infty}^{+\infty} p(u,t) du$ and $u$, as before, is a shorthand for $(u_1 \dots u_N)$.

If the dependence on each of the $N$ independent variables $u_j$ is irreducible, 
the average of $u_j$ depends on the full N-body PDF $p_N(u_1 \dots u_N)$, with no room
for marginalization.
But this is rarely the
case in classical physics.
For instance, if $K=\sum_{j=1}^N u_j^2$ is the total kinetic energy of the fluid,
its average depends only on the one-point distribution
\begin{equation}
\langle K\rangle (t) = \sum_{j=1}^N \int u_j^2 p_N(u) du = \sum_{i=1}^N \int u_i^2 p_1(u_i) du_i,
\end{equation}  
where $p_1(u_i)$ results from integrating out all 
variables $u_1$ to $u_N$, but $u_i$.
Likewise, if all we need is the value of the average velocity field at the space 
slice $x_i$, the 1-point PDF will suffice
\begin{equation}
\langle u_i \rangle (t) =  \int u_i p_1(u_i,t) du_i.
\end{equation}  
By the same token, two-point observables require two-point PDFs, and
so on, at all higher orders. That said, we proceed to estimate the computational viability 
of the marginalization procedure.

The count goes as follows.
The N-body PDF associated with a discrete grid with $G$ lattice sites, 
each hosting $F$ fields discretized over a set of $n$ discrete values,
takes on $(GF)^n$ discrete values.
The number of qubits to represent the fully N-body discrete PDF is then
\begin{equation}
\label{QG}
q = GF log_2 n.
\end{equation}
Given that $G$ is in the order of 
many billions for present-day supercomputer 
simulations, this requires multi-billions logical 
bits, a number which appears to be totally unrealistic in any
foreseeable future. Fortunately, marginalization presents a much more optimistic picture.

If each field on a discrete grid with $G$ lattices sites is connected to
$z<<G$ neighbours, the lowest order irreducible marginal
is of order $M=zF$ and the qubit count now reduces to
\begin{equation}
\label{QZF}
q = zF log_2 n.
\end{equation}
This is still very demanding but vastly simpler than (\ref{QG})
since for the three-dimensional Navier-Stokes equations, the
parameter $zF$ is on the order of tens.  This shows that marginalization stands good chances to circumvent
the dimensional curse on a quantum computer.
In the succeeding sections we provide a more quantitative 
assessment in this direction. A useful reference in this context is \cite{Daley2019}.

\section{Practical Examples}

We begin by inspecting the Liouville formulation
for the case of the Burgers equation, describing a one-dimensional
pressure-free fluid.  Since we consider the noiseless Burgers
equation (whose noisy version is the famous Kardar-Parisi-Zhang
equation), we set the diffusion to zero below.

\subsection{The Burgers-Liouville Equation}

The Burgers equation describing one-dimensional 
pressureless fluids, reads
as 
\begin{equation}
\partial_t u + u \partial_x u = \nu \partial_{xx} u,
\end{equation}
where $\nu$ is the kinematic viscosity.

A simple center-finite difference scheme gives
\begin{equation}
\dot u_j = - u_j (u_{j+1}-u_{j-1}) + \nu (u_{j+1}-2 u_j + u_{j-1}) 
\equiv \sum_{k=-1}^1 B_{j,k}(u_j) u_{j+k} \equiv f_j(u), 
\end{equation}
where the space step is made unity for simplicity and
$B_{jk}$, $j=1,N$, $k=-1,0,1$  is the (nonlinear) ``Burgers" matrix.
In explicit form, we have:
\begin{equation}
B_{j,j-1} = \nu -\frac{1}{4} u_j,\;
B_{j,j} =   -2 \nu -\frac{1}{4} u_j,\;
B_{j,j+1} = \nu +\frac{1}{4} u_j. 
\end{equation}

The N-point Burgers-Liouville equation takes the following form:
\begin{equation}
\partial_t p_N + \sum_{j=1}^N
\partial_{u_j} [\sum_{k=-1}^1 B_{j,k} u_{j+k}] p_N. 
\end{equation}\
Here $p_N \equiv p(u_1 \dots u_N,t)$ is the N-body PDF associated
with the spatial grid of $G=N$ points.
\begin{figure}
\centering
\includegraphics[scale=0.3]{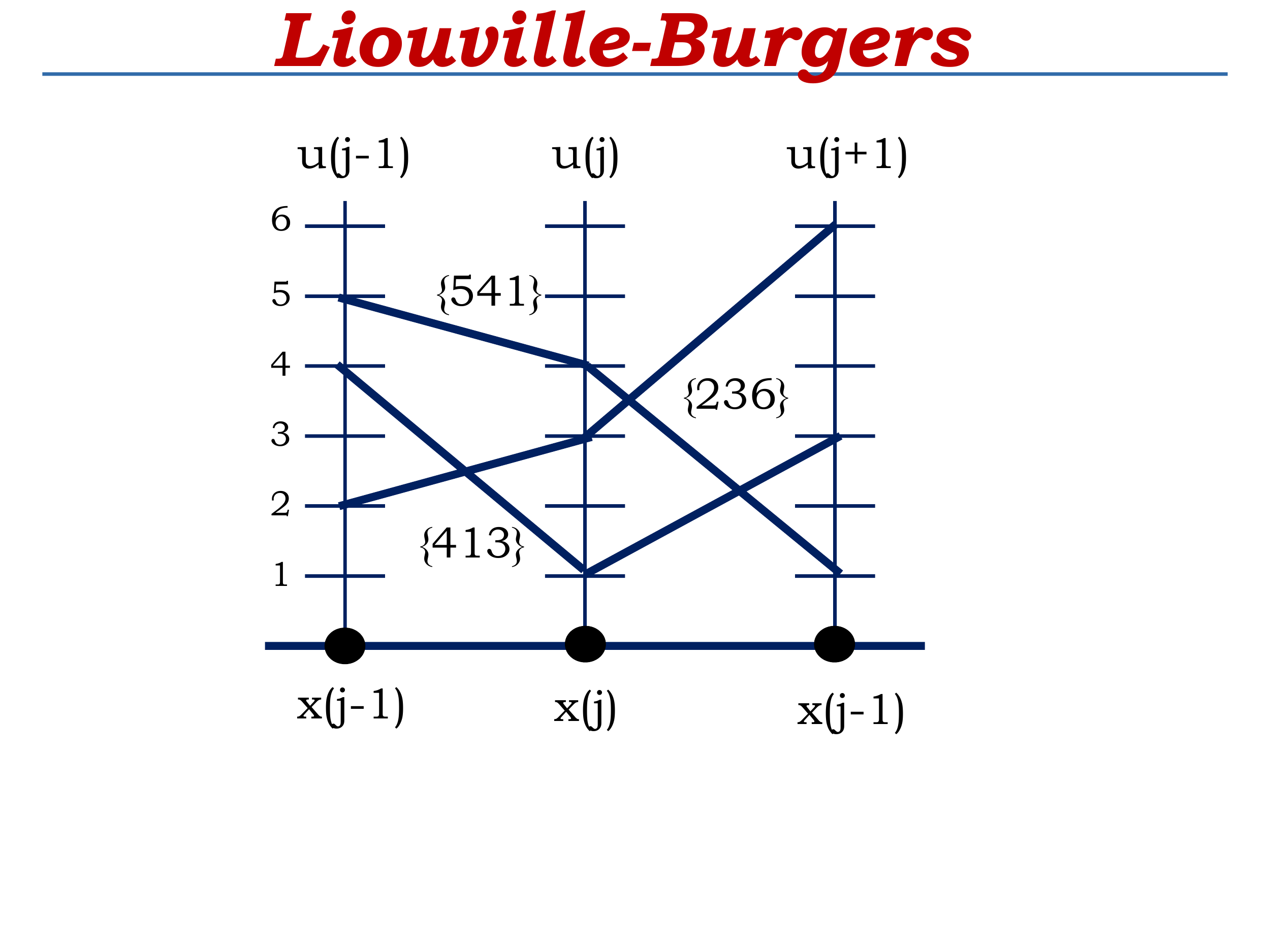}
\caption{Geometrical interpretation of the Liouville-Burgers equation.
Each of the three independent variables $u_j,u_{j-1},u_j$ takes up to $n=6$ 
values, hence the discrete Liouville-Burgers equation takes values on a set
of three-points paths labeled by three integers $n_{j-1},n_{j},n_{j-i}$,
each varying between $1$ and $6$. The figure reports the paths 
$\lbrace 413 \rbrace$, $\lbrace 236 \rbrace$ and $\lbrace 541 \rbrace$.  
}
\end{figure}

Since ${\dot u}_j$ depends on the triplet $(u_{j-1},u_j,u_{j+1})$,the
lowest order irreducible marginal is the three-point PDF $p_3$, which is
defined by integrating out all independent variables 
but three, $u_{j-1},u_j,u_{j+1}$. 
This gives:
\begin{equation}
p_3(u_{j-1},u_j,u_{j+1}) = \int_{\infty}^{+\infty} 
p(u_1 \dots u_N) du_1 \dots du_{j-2} du_{j+2} \dots du_N. 
\end{equation}
The corresponding 3-point kinetic equation takes the form: 
\begin{equation}
\partial_t p_3 + 
\partial_{u_{j-1}} [\sum_{k=-1}^1 B_{j-1,k} U_{j-1+k}] p_3 +
\partial_{u_{j}}   [\sum_{k=-1}^1 B_{j,k}   U_{j+k}]   p_3 +
\partial_{u_{j+1}} [\sum_{k=-1}^1 B_{j+1,k} U_{j+1+k}] p_3 = 0.
\end{equation}
Here, by periodicity, $j-2=j+1$ and $j+2=j-1$.

In the above we have defined
\begin{equation}
U_j(t) = \int u_j p(u,t) du_1 \dots du_{j-2} du_{j+1} du_N ,
\end{equation}
which is a generally unknown function of $u_{j-1},u_j,u_{j+1}$.
Hence a suitable expression for $U_j$ versus $u_j$
needs to be worked out, which is the usual closure problem.
In the following, we shall proceed on the assumption that such
a closure can be worked out.

With reference to the expression (\ref{QZF}), we have
$z=3$ and $F=1$, hence the corresponding qubit count gives 
\begin{equation}
q = 3 \; log_2 n.
\end{equation}
Current quantum computers feature up to $q \sim 500$ {\it nominal} qubits 
\cite{IBM}, implying that one can reach up to $n \sim 2^{500/3}$, far beyond 
any practical resolution needed.
For a reasonable resolution, say $n \sim 1000 \sim 2^{10}$, we obtain $q \sim 30$, which
appears viable once noise and decoherence are tamed. 

\subsection{Example 2: The Navier-Stokes Liouville Equation}

The Navier-Stokes governing the motion of compressible, dissipative fluids
read as follows
\begin{eqnarray}
\partial_t \rho + \partial_a (\rho u_a) = 0,\\
\partial_t (\rho u_a + \partial_b (\rho u_a u_b + P \delta_{ab} - \sigma_{ab}) = 0,
\end{eqnarray}
where $\rho$ is the density, $u_a$, ($a=x,y,z$) the flow velocity,
$P=P(\rho)$ the fluid pressure and $\sigma_{ab}$ the dissipative tensor.

With reference to the expression (\ref{QZF}), we now have
$F=4$ (density and three velocity components) and $z=7$ (each grid site
connected to six nearest neighbors), hence the corresponding qubit count gives 
\begin{equation}
q = 28 \; log_2 n .
\end{equation}
With $n=10^3$, we have $q=280$, much larger than for Burgers, but still
within the nominal capabilities of current quantum hardware \cite{IBM}. 
Different representations or altogether different formulations, such as as 
lattice Boltzmann \cite{LB} methods, may lead to more favourable scalings, this being a topic of interest
for future research.

The above considerations reveal the many issues 
generally associated with quantum computing, particularly noise and decoherence.
In the following, we briefly comment on a further issue which is peculiar to
quantum simulations in real time, namely time marching. 

\section{Sketch of the Quantum Algorithm}

As mentioned above, a number of different quantum computing strategies
have been proposed in the recent past to simulate fluid 
problems on quantum computers.
While all of these methods need to handle the nonlinearity issue,
there is no such need in our case, since the problem is linear from scratch; one can proceed by resorting to quantum linear 
solver algorithm \cite{QLAS}, as detailed below.

We start by writing the 3-point Liouville equation in the explicit conservative form
\begin{equation}
\partial_t p + \partial_u (Up) + \partial_v (Vp) + \partial_w (Wp) = 0,
\end{equation}
where we have set $u \equiv u_{j-1}$, $v \equiv u_j$ and $w \equiv u_{j+1}$
and $(U,V,W)$ are three supposedly known functions of $(u,v,w)$.
Upon discretizing the three-dimensional functional space $(u,v,w)$, 
$(u_i,v_j,w_k)$, $i,j,k=1,n$,
we obtain a set of $G=n^3$ ODE's of the form
\begin{equation}
\dot p_{ijk} + U_{ii',jk}p_{i',jk} + V_{i,jj',k} p_{ij'k} + W_{ij,kk'}p_{ij,k'} = 0,
\end{equation}
where $U,V,W$ are the discrete matrices associated with the three flux terms.

A simple Euler forward time marching delivers
\begin{equation}
p^{t+1} - (1-Ldt)p^t = 0,
\end{equation}
with the initial condition $p^0 = p_0$, $t$ labelling discrete time,
and $p_0$ being the initial condition, all spatial indices being suppressed
for simplicity and $L$ denotes the sum of the matrices, $L=U+V+W$.
The above relations deliver a linear system for the 
unknown $p ={p^{(0)},p^{(1)} \dots p^{(M)}}$, each component being 
an array of dimension $n^{3}$.
In explicit form,
\begin{eqnarray}
p^{(0)}                  = p_0\\
p^{(m+1)}-(1-Ldt)p^{(m)} = 0\\
m=0,\dots M-1,
\end{eqnarray}
where $M=T/dt$ is the number of time slices.
This is a linear system $Ap=q$, with a sparse lower-triangular matrix 
structure and $q$ features just a single entry, $p_0$. 
As a result, it is readily solved as a causal sequence of matrix-vector
products. It is still formally a linear system;  
as such, it can be handled by quantum linear 
solver algorithms, as shown in \cite{CHILDS} for the case of the Burgers 
equation on the order of $O(10-100)$ grid points.

The computational quantum complexity is given by
\begin{equation}
\mathcal{C}_q \propto s \phi T^2 Polylog (G, 1/\epsilon),
\end{equation}
where $s$ is the sparsity of the matrix, $\phi$ is the fidelity of the
initial condition, $T$ the time-span, $G$ the grid size, and $\epsilon$
is the error tolerance.

A classical explicit algorithm would scale instead like
\begin{equation}
\mathcal{C}_c \propto s T G,
\end{equation}
showing that the main advantage is the $poly log (G)$ factor, partly 
reabsorbed by the $T^2$ dependence on the time-span.
Since $T \sim G^{1/3}$, a significant quantum speed-up can be expected.

\section{Comparison with the Dynamic Approach}

In the beginning of this paper, we have cautioned the reader about the practical
unviability of the Liouville approach on classical computers, as the problem
occupies a space with about thirty dimensions.
Hence, on classical computers there is no option but solving repeated realizations
of the fluid equations.
It is therefore of interest to assess the cost of the dynamic approach 
on quantum computers.
Current approach provide log scaling in the number of grid points and typically
a quadratic scaling in time.
With such a scaling, running a billion grid points ($10^{9}$) over a million
time steps, would take about $10^{21}$ operations, hence that many dynamic
degrees of freedom. By running an ensemble of $1000$ simulations, say, this
comes to $10^{24}$, namely about $80$ qubits, much less
than for the Liouville equation.  
The problem though is that time-marching on a quantum computers involves
the reconstruction of the full quantum state at each time step (by virtue of the no-cloning
theorem), an operation which scales exponentially with the number of qubits.
For the case of a billion grid points, namely 30 qubits, this adds a 
a dramatic slowdown, which is a major issue in quantum computing \cite{CFDVQC}.

It thus appears that ensemble simulations of fluid flows 
on quantum computers are best performed via the Liouville approach, provided
(i) a sensible closure can be worked out, and (ii) hundreds
of reliable logical qubits can be used.
Finally, we remark that a similar statement applies to 
basically any nonlinear field theory.

\section{Summary}

Summarizing, we have assessed the viability of the functional Liouville
formulation for ensemble simulations of fluid flows on quantum computers.
The present analysis refers to a blue-sky scenario whereby a quantum 
algorithm capable of logarithmic scaling with the number of dynamic 
degrees of freedom is available {\it and} running on {\it ideal}
quantum computers, with no appreciable decoherence and/or noise problems.

In actual practice, quantum computing ensemble simulations of the
Navier-Stokes equations demand hundreds of noiseless {\it logical} qubits.
Given that current quantum computing typically works only up to a few
logical qubits, say of the order ten, the target appears to be in the future.  
This is no invitation to surrender, but just a realistic appraisal to be
contrasted with the current (mostly commercial) hype around quantum computing
(for a highly informed assessment, see \cite{HYPE}).

\section{Acknowledgements}

The authors have benefited from valuable discussions 
with many colleagues, particularly
S.S. Bharadwaj, 
D. Buaria,
P. Coveney, N. Defenu, A. Di Meglio,
M. Grossi, A. Mezzacapo, 
S. Ruffo, A. Solfanelli and T. Weaving.
S.S. acknowledges financial support form the Italian National
Centre for HPC, Big Data and Quantum Computing (CN00000013).

\end{document}